\documentstyle[epsf,twocolumn]{mn}

\date{submitted to MNRAS}

\title[Minkowski Functionals of Abell/ACO Clusters]
{Minkowski Functionals of Abell/ACO Clusters}

\author[M. Kerscher et al.]
{M.~Kerscher$^1$, J. Schmalzing$^1$, J.~Retzlaff$^2$, 
S.~Borgani$^{3,4}$, T.~Buchert$^1$,\vspace*{1ex}\\
\LARGE S.~Gottl\"ober$^2$, 
V.~M\"uller$^2$,  M.~Plionis$^{4,5}$ \& H.~Wagner$^1$ \vspace*{1ex}\\ 
$^1$Sektion Physik, Ludwig--Maximilians--Universit\"at,
Theresienstr.~37, D--80333 M\"unchen, Germany\\
$^2$Astrophysikalisches Institut Potsdam, An der Sternwarte 16,
D--14482 Potsdam, Germany\\ 
$^3$INFN Sezione di Perugia, c/o Dipartimento di Fisica
dell'Universit\`a, via A. Pascoli, I--06100 Perugia, Italy \\
$^4$SISSA -- International School for Advanced Studies, via Beirut
2--4, I--34013 Trieste, Italy \\
$^5$National Observatory of Athens, Lofos Nimfon, Thesio, 18110
Athens, Greece }

\newcommand{\mincir}{\raise -2.truept\hbox{\rlap{\hbox{$\sim$}}\raise5.truept
\hbox{$<$}\ }}
\newcommand{\magcir}{\raise -2.truept\hbox{\rlap{\hbox{$\sim$}}\raise5.truept
\hbox{$>$}\ }}
\newcommand{\minmag}{\raise-2.truept\hbox{\rlap{\hbox{$<$}}\raise 6.truept\hbox
{$>$}\ }}

\newcommand{\hm}{\,h^{-1}{\rm Mpc}}
\newcommand{\vel}{\,{\rm km\,s^{-1}}}

\def\hMpc{\ifmmode{h^{-1}{\rm Mpc}}\else{$h^{-1}$Mpc}\fi}
\def\R{{\rm I\kern-0.16em{}R}}
\def\etal{ et al.\ }
\renewcommand{\baselinestretch}{1.0}
\begin{document}

\maketitle

\begin{abstract}
We determine the Minkowski functionals for a sample of Abell/ACO
clusters, 401 with measured and 16 with estimated redshifts.  The four
Minkowski functionals (including the void probability function and the
mean genus) deliver a global description of the spatial distribution
of clusters on scales from $10$ to $60\hMpc$ with a clear geometric
interpretation. Comparisons with mock catalogues of N--body simulations
using different variants of the CDM model demonstrate the
discriminative power of the description. The standard CDM model and
the model with tilted perturbation spectrum cannot generate the
Minkowski functionals of the cluster data, while a model with a
cosmological constant and a model with breaking of the scale invariance of
perturbations (BSI) yield compatible results.
\end{abstract}


\section{Introduction}

The distribution of galaxy clusters has been used for a long time as a
useful tracer of the large--scale structure of the Universe and as a
constraint for cosmological models.  To present knowledge, clusters of
galaxies represent the largest gravitationally bound entities in the
hierarchy of cosmic structures.  In addition, their amplified
clustering, with respect to that of galaxies, can be reliably measured
on large scales ($> 20\hm$) where the gravitational evolution still
keeps track of the initial fluctuation spectrum.

The first quantitative analyses of the cluster distribution date back
to the pioneering works by Bahcall \& Soneira (1983) and Klypin \&
Kopylov (1983) and were based on the estimate of the cluster 2--point
correlation function, $\xi_{cc}(r)$.  Although dealing with a rather
limited number of objects, these analyses clearly showed that Abell
(1958) clusters have a clustering scale length
$r_0$, defined by
$\xi_{cc}(r_0)=1$, which is much larger than that of
galaxies. Considerable observational effort has been made to improve
the large--scale mapping of the cluster distribution.  Analyses based
on the extension of the Abell sample to the southern hemisphere
(Abell, Corwin \& Olowin 1989, hereafter ACO) and on progressively
larger redshift compilations of Abell and ACO clusters (e.g. Postman,
Geller \& Huchra 1992) basically confirmed the original result that
$r_0\simeq 20\hMpc$ (see also Cappi \& Maurogordato 1992; Plionis,
Valdarnini \& Jing 1992; Peacock \& West 1992).  On the other hand,
new samples based on more objective cluster identification criteria,
both in the optical (Dalton\etal 1994; Collins\etal 1994) and in the
X--ray (e.g. Nichol, Briel \& Henry 1994; Romer\etal 1994) bands,
revealed a smaller length scale, $r_0\simeq13$\ldots16$\hMpc$.  This
difference might be ascribed to richness contamination from projection
effects of the Abell/ACO samples (Sutherland 1988, see also Jing,
Plionis \& Valdarnini 1992) or to the richness--clustering dependence
(Bahcall \& West 1992). In any case, the standard CDM model fails to
predict the large range of positive cluster correlation function
($\xi(r)>0$ up to scales of $70\hMpc$, see White\etal 1987,
Olivier\etal 1993).

Although several further analyses have been devoted to compare data on
the cluster distribution to the predictions of a variety of DM models
(e.g. Bahcall \& Cen 1992; Dalton\etal 1994; Borgani\etal 1995), most
of them have been mainly based on low--order statistics. A more
complete description of the cluster distribution requires the use of
higher--order statistics, which are capable of capturing global
aspects of the clustering. One example is the genus statistics applied
to the cluster data by Rhoads, Gott \& Postman (1994)
and by Pearson et al. (1996).

A direct measurement of properties of a general point distribution can
be made using Minkowski functionals (Mecke, Buchert \& Wagner
1994). These measures provide a morphological (geometrical and
topological) description of the point distribution and comprise
information about correlation functions of arbitrary
order. Application to clusters of galaxies is especially promising
since clusters bear direct contact to the formation of the largest
structures in the universe. As geometrical characteristics of
structure, the Minkowski functionals combine both the advantage of
intuitive interpretation of their meaning and the advantage of
delivering a quantitative measure. The aim of this paper is to apply
the Minkowski functional statistics in order to compare an extended
redshift sample of Abell/ACO clusters with mock catalogues obtained
from large PM N--body simulations, starting from different models for
the initial power spectrum. To this purpose, we extract artificial
cluster samples from the simulations, thereby reproducing the main
characteristics of the observational data set (cluster number density,
boundary geometry, selection functions for galactic absorption and
luminosity effects).

Simulations are run for the Standard Cold Dark Matter (SCDM) model, a
`Tilted' CDM model (TCDM) with $\nu=0.9$ for the post--inflationary
spectral index, a low--density CDM model with $\Omega_0=0.35$
and a cosmological constant ($\Lambda$CDM), and a double inflation
model with Broken Scale Invariance (BSI) of the primordial
perturbation spectrum. 
It is generally accepted that the SCDM spectrum fails to reproduce the
cluster correlation amplitude and, once normalized on large scales to
match the CMB anisotropies, it largely overproduces clusters. These
shortcomings are interpreted as the consequence of the wrong shape of
the SCDM spectrum (see, e.g., Borgani et al. 1996; Mo, Jing \& White
1996). As a possible remedy, $\Lambda$CDM modifications of the SCDM
model suitably change the spectrum shape at the cost of introducing
one additional parameter in comparison to SCDM, while the TCDM model
and the BSI model change the shape of the power spectrum by
introducing one and two additional parameters, respectively.

A brief description of the observational data is provided in Section
2. In Section 3 we present the cosmological models on which
simulations are based and sketch the procedure for cluster selection
and mock catalogue construction.  We define and discuss in Section 4
the properties of Minkowski functionals, relations to other
statistics, the method of analysis and its results.  In Section 5 we
summarize and draw our conclusions.

\section{Observational data }

We consider an extended redshift sample of Abell and ACO clusters with
richness $R\ge 0$ (Abell 1958; Abell, Corwin \& Olowin 1989).  In the
following we will provide only a brief description of this sample
since more details are given in Borgani\etal (1996; see also Plionis
\& Valdarnini 1995).  The northern (Abell) part of the sample, with
declination $\delta \ge -17^{\circ}$, is defined by those clusters
that have measured redshift $z\le 0.1$, while the southern ACO part,
with $\delta < -17^{\circ}$, is defined by those clusters with
$m_{10}< 17$, where $m_{10}$ is the magnitude of the tenth brightest
cluster galaxy in the magnitude system corrected according to Plionis
\& Valdarnini (1991).

The effect of galactic absorption is modelled according to the
standard cosecant dependence on the galactic latitude $b$,
\begin{equation}\label{eq:obs1}
P(|b|)=10^{\alpha\left(1-\csc|b|\right)},
\end{equation}
with $\alpha \approx 0.3$ for the Abell sample (Bahcall \& Soneira
1983; Postman\etal 1989) and $\alpha \approx 0.2$ for the ACO sample
(Batuski\etal 1989).  In order to limit the effects of galactic
absorption we only use clusters with \mbox{$|b|\ge 30^{\circ}$}.

The cluster--redshift selection function $P(z)$ is determined by
fitting the cluster density as a function of~$z$:
\begin{eqnarray}
P(z)~=~\left\{ \begin{array}{ll}
 1               & \mbox{if $z\le z_c$,} \\
 A\,\exp(-z/z_o) & \mbox{if $z>z_c$,}
\end{array}
\right.
\label{eq:pz}
\end{eqnarray}
where $A=\exp{(z_c/z_o)}$, and $z_c$ is the redshift below which the
spatial density of clusters remains constant (volume--limited sample).
We find $z_c\approx 0.078$, $z_o\approx 0.012$ and $z_c\approx 0.068$,
$z_o\approx 0.014$ for Abell and ACO samples, respectively.  Since the
exponential decrease of $P(z)$ can introduce considerable shot noise
errors at large redshifts, we prefer to limit our analysis to $r_{\rm
max}=240\hMpc$, where the catalogue is approximately volume limited.

There are in total 417 Abell/ACO clusters fulfilling the above
criteria: 262 Abell clusters with measured redshifts, and 155 ACO
clusters, 16 of which have $z$ estimated from the $m_{10}$--$z$
relation calibrated by Plionis \& Valdarnini (1991).  These numbers
correspond to $\langle n\rangle_{\rm Abell}= (1.6 \pm 0.25) \times
10^{-5} \;(\hMpc)^{-3}$ and $\langle n\rangle_{\rm ACO}= (2.3 \pm 0.3)
\times 10^{-5} \;(\hMpc)^{-3}$, for the Abell and ACO cluster number
densities, respectively, once corrected for galactic absorbtion
according to eq. (\ref{eq:obs1}). The density difference is mostly
spurious, due to the higher sensitivity of the IIIa--J emulsion plates
on which the ACO survey is based. The above density values correspond
to average cluster separations of $\langle d_{\rm Abell}\rangle \simeq
40\;\hMpc$ and $\langle d_{\rm ACO}\rangle \simeq 35\;\hMpc$.

\section{Mock samples from N--body simulations}\label{mock-sect}

\subsection{Cosmological models}

We simulated the evolution of large--scale structure in four different
types of spatially flat Cold Dark Matter models.  They are listed
below.

\begin{description}
\item[(a)] The Standard CDM (SCDM) model with \mbox{$\Omega_0=1$} and
$h=0.5$ for the Hubble constant in units of $100\vel$ Mpc$^{-1}$,
which we take as a standard of reference.
\item[(b)] A tilted CDM (TCDM) model, with $\nu=0.9$ for the
post--inflationary spectral index as arising, for instance, from power
law inflation.
\item[(c)] A low--density CDM ($\Lambda$CDM) model, with
\mbox{$\Omega_0=0.35$}, $h=0.7$ and spatial flatness restored by a
cosmological constant term $\Omega_\Lambda=0.65$.
\item[(d)] A Broken Scale--Invariance (BSI) CDM model, which exhibits
a step--like primordial spectrum due to a double--inflationary
scenario. The modification of the power spectrum is specified by two
parameters, the step location at \mbox{$k_{\rm{break}}^{-1}=1.5\hm$}
and its relative height $\Delta=3$.  With this parameter choice, the
model provides a good fit to a variety of observational data
(Gottl\"ober, M\"ucket \& Starobinsky 1994, Amendola\etal 1995,
Kates\etal 1995, Ghigna\etal 1996).
\end{description}

For the CDM model we use the transfer function as parameterized by
Bardeen\etal (1986), which assumes a vanishing baryon contribution.
Although the presence of baryons is relevant as far as small--scale
properties ($\mincir 5\hm$) are concerned, it has negligible influence
on the large--scale clustering ($\magcir 10\hm$), which we want to
investigate using galaxy clusters. We summarize the main model parameters
in Table~\ref{model-param}.  The power spectra are normalized
according to the two year COBE--DMR 53 and 90 GHz galactic sky maps
following the prescription of 
G\'orski\etal (1994) for the $\Omega=1$ models and of 
Stompor, G\'orski \& Banday (1995) for the $\Lambda$CDM model.  

We did not take into account any possible gravitational wave
contributions for the BSI and TCDM model. We have also done our
analysis for the one year COBE normalization and verified that the
statistics of the cluster distribution are insensitive to this
reduction of the spectrum amplitude (see also Croft \& Efstathiou
1994; Borgani\etal 1995). Therefore, any uncertainty either in our
normalization procedure or in the measured level of CMB temperature
anisotropy has no effect on the final results of the analysis. In
particular, our results will not change when adopting a power spectrum
normalization compatible with the 4 year COBE data (see, e.g., Bennett
et al. 1996). This normalization would imply a reduction of the
fluctuation amplitude $\sigma_8$ by about 11~\% for all the
models under consideration.

\begin{figure}
\begin{center}
\epsfxsize=7.8cm
\begin{minipage}{\epsfxsize}
\epsffile{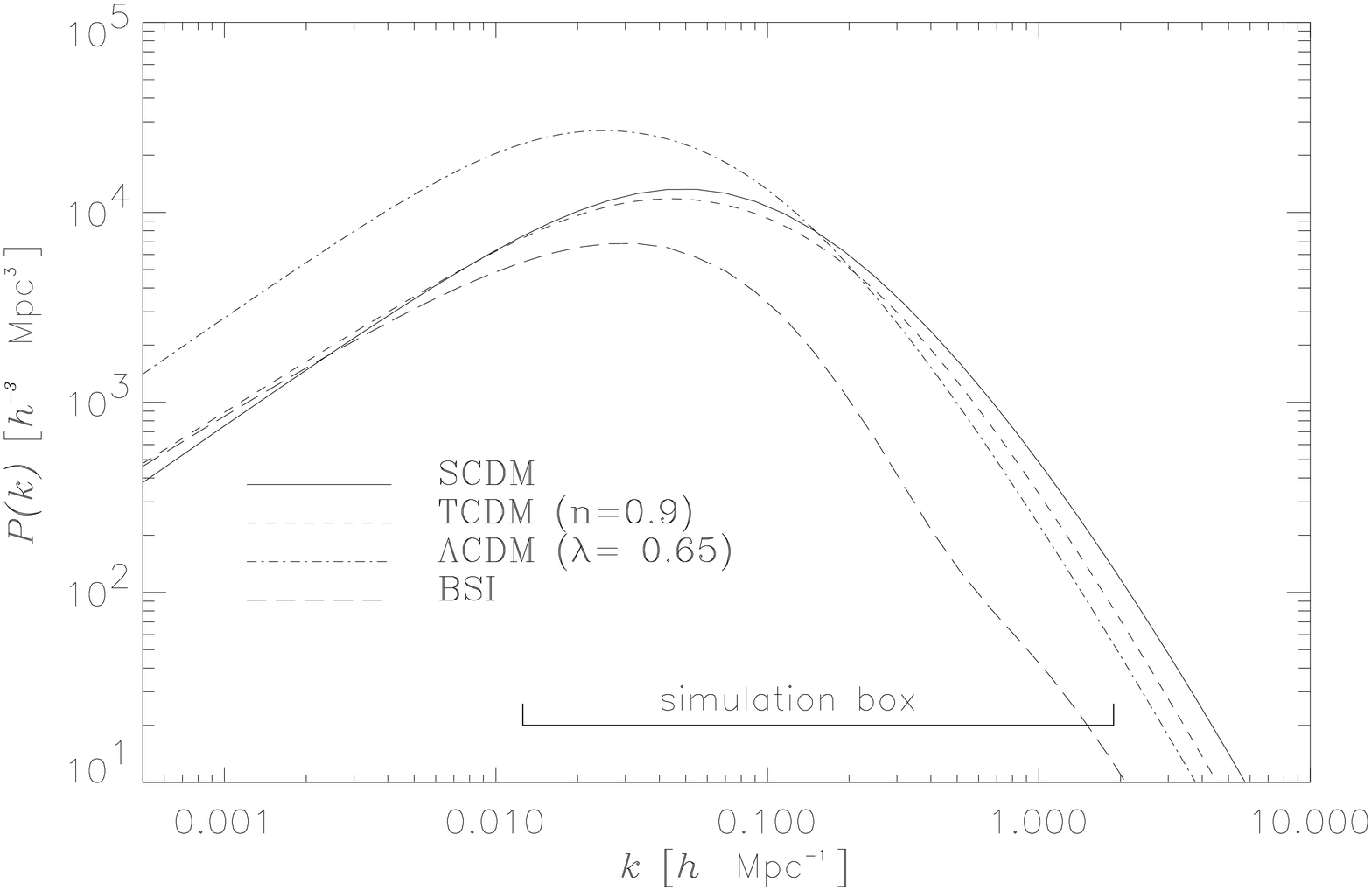}
\end{minipage}
\end{center}
\caption{\label{powerspectrum}
The SCDM, TCDM, $\Lambda$CDM and BSI power spectra. The bar selects the
wavenumber range corresponding to the adopted box size of $500\hm$.}
\end{figure}

The spectra we used in our simulations are shown in
Figure~\ref{powerspectrum}, where the horizontal bar selects the part
of the spectrum between the inverse box size and the Nyquist frequency
which is realized in the simulation.

We evolve the initial density field starting from redshift $z=25$
until the present epoch employing a standard PM N--body scheme with
$N_{\rm{p}}=300^3$ particles ($m_{par} = 1.3 \times 10^{12}
h^{-1}{\rm{M_\odot}}$) and $N_{\rm{g}}=600^3$ grid cells in a
simulation box of $L=500\hm$ comoving length a side.  This provides a
spatial resolution of $\simeq 1.7\hMpc$ (two cells).  We suppose that
the simulation box is large enough to contain all fluctuation modes
which contribute to the large--scale cluster clumping.  In order to
account for the effect of statistical variance, we carried out
simulations for four random realizations for SCDM and three for
$\Lambda$CDM, respectively, while one realization was done for
TCDM and BSI.

\subsection{Mock cluster samples}

We identify clusters from the final particle distribution by applying
an iterative procedure to the high peaks of the density field
reconstructed on the $600^3$ mesh points (see also Klypin \& Rhee
1994).  After assigning the density field on the grid through a
cloud--in--cell interpolating scheme, we identify those points which
correspond to local density maxima.  Afterwards, we center a sphere of
radius $1.5\hm$, corresponding to the Abell radius, on each of these
local maxima and compute the center of mass position for all the
particles falling within that sphere.  This position is used as the
new cluster center and the procedure is repeated until convergence.
We find that in general only a few (about 5) iterations are required for
the cluster coordinates and masses to converge to their final values.

\begin{table}
\begin{center}
\caption{\label{model-param}Summary of model parameters}
\begin{tabular}{cccccc} \hline
Model        & $h$ & $\Omega_0$ & $\Omega_\Lambda$ & $\nu$ & $\sigma_8$ \\
\hline
SCDM          & 0.5 & 1                 & 0                & 1     & 1.37 \\
TCDM         & 0.5 & 1                 & 0                & 0.9   & 1.25 \\
$\Lambda$CDM & 0.7 & 0.35              & 0.65             & 1     & 1.30 \\
BSI          & 0.5 & 1                 & 0                & --    & 0.60 \\
\hline
\end{tabular}
\end{center}
\end{table}

From the resulting list of candidate clusters, we select the $N_{cl}$
most massive objects and identify them with Abell/ACO clusters. By
definition, $N_{cl}=(L/d_{cl})^3$ is the expected number of clusters
within the simulation box, having mean separation $d_{cl}$.

In order to check the robustness of our cluster identification scheme,
we also based their identification on the friend--of--friend
algorithm, with linking length $b=0.2$ times the mean particle
separation (e.g. Frenk\etal 1988).  This value of $b$ defines groups
bounded by an isodensity surface of about 125 times the average
density and, therefore, to an average internal overdensity of about
180 if an isothermal density profile is assumed (e.g. Lacey \& Cole
1994).  Such an overdensity is very close to the value expected for a
virialized structure resulting from the spherical top--hat collapse
(e.g. Peebles 1980). We will show elsewhere that the cluster
distributions obtained with these two procedures are very similar not
only in a statistical sense, but also as far as the point--to--point
comparison of the cluster positions is concerned.

\begin{figure}
\begin{center}
\epsfxsize=5cm
\begin{minipage}{\epsfxsize}
\epsffile{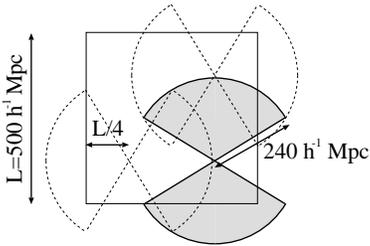}
\end{minipage}
\end{center}
\caption{\label{mock-geometry}
Projected positions of three mock samples in the simulation box,
showing their overlap.}
\end{figure}

After generating the cluster distribution in the simulation box, we
extract mock samples which reproduce the same observational features
as the real Abell/ACO sample.  In each box we locate 8 observers along
the main diagonal axes, each having a distance of $L/4=125\hm$ from
the three closest faces.  First we include all clusters up to a
maximum distance of $240\hm$ in each mock sample (see Figure
{}\ref{mock-geometry}, the simulation box has periodic boundaries). We
then randomly sample them to get a density distribution reproducing
the observational selection functions for galactic absorption and
redshift extinction (see also Borgani\etal 1996).
In order to minimize the overlap between mock samples, the coordinate
systems for two adjacent observers are chosen so that the
corresponding galactic planes are orthogonal to each other. Even with
this choice, it turns out that different mock samples involve
overlapping volumes and, therefore, they cannot be considered as
completely independent.

Minkowski functionals are sensitive to the number density.  In the
case of a Poisson process one may easily derive scaling relations with
the number density and radius using the analytical results of Mecke \&
Wagner (1991) and the homogeneity property of the Minkowski
functionals, $M_\mu(\lambda A_r)=\lambda^{d-\mu}M_\mu(A_r)$,
($\mu=0,\dots,3)$ with a positive real scaling factor $\lambda$ (see
Section \ref{min-sect} for the definition of the functionals $M_\mu$
of $A_r$ produced by balls $B_r$ of radius $r$ around each cluster).
For generic distributions no scaling relations are available since the
scaling properties depend, even without correlations, on the
dimensionality of the support of the point process. Therefore, care
has to be taken to reproduce in the simulated samples the correct
number of Abell and ACO clusters separately.  To this purpose, after
generating a mock sample, we randomly degrade the cluster number
density in the Abell part until the number density reaches a fraction
$\langle n\rangle_{\rm Abell}/\langle n\rangle_{\rm ACO}$ of the
number density of the ACO part.  Since the overall number density is
fixed in the whole simulation box, different samples may contain
different numbers of clusters, with fluctuations around $10\%$ and
deviations of single samples as large as $30\%$.  Instead of forcing
all the samples to have the same number of clusters as the real one,
we prefer to maintain such fluctuations and consider them as an effect
of cosmic variance.  We only take care that, after averaging over the
eight observers in each box, the resulting average cluster number per
sample reproduces the observational one. We find that choosing a mean
cluster separation $d_{cl}=36\hm$, which roughly corresponds to the
average separation of ACO clusters, always produces an average number
of Abell and ACO clusters per sample which differs by $\mincir 5\%$
from the corresponding numbers of the observational case.

As we will see, uncertainties in the analysis of the Minkowski
functionals are largely dominated by the observer--to--observer scatter
within the same box, which is due to the fluctuations in the cluster number
density. The box--to--box variance of different simulations, which
accounts for the sampling of independent patches of the Universe is
significantly smaller.

\section{Analysis using Minkowski functionals}\label{min-sect}

\subsection{Definition and general properties}

\begin{table}
\caption{\label{min-3d-geom}
The Minkowski functionals in three--dimensional space expressed in
terms of the corresponding geometric quantities.}
\begin{center}
\begin{tabular}{|cl|c|c|}
\hline
    & geometric quantity      & $\mu$ & $M_\mu$      \\ 
\hline						     \\ %
$V$ & volume                  & 0     & $V$          \\ 
$A$ & surface                 & 1     & $A/8$        \\ 
$H$ & integral mean curvature & 2     & $H/2\pi^2$   \\ 
$\chi$ & Euler characteristic & 3     & $3\chi/4\pi$ \\ 
\hline
\end{tabular}
\end{center}
\end{table}

\begin{figure}
\epsfxsize=2in
\begin{center}
\begin{minipage}{\epsfxsize}\epsffile{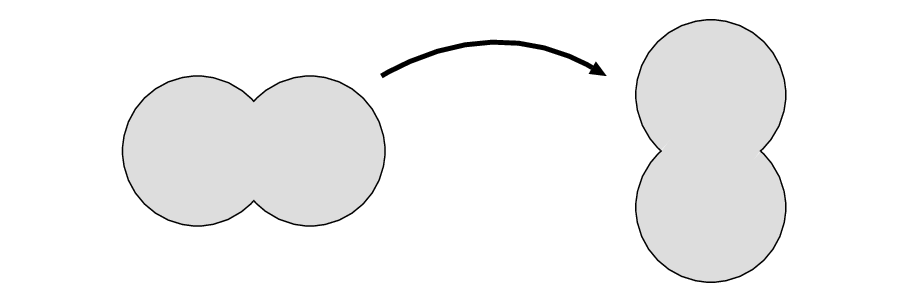}\end{minipage} \\ 
$M_\mu(A) = M_\mu(gA)$						 \\
\begin{minipage}{\epsfxsize}\epsffile{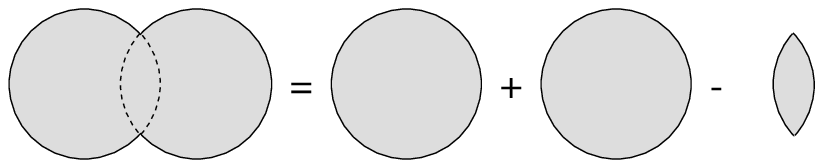}\end{minipage} \\
$M_\mu(B_1\cup B_2) = M_\mu(B_1)+M_\mu(B_2)-M_\mu(B_1\cap B_2)$	 \\
\begin{minipage}{\epsfxsize}\epsffile{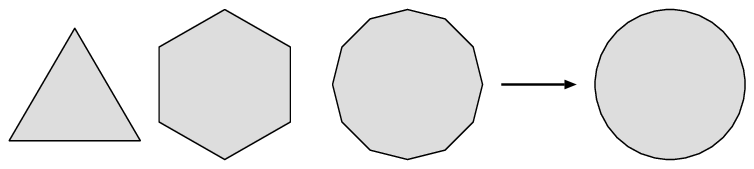}\end{minipage} \\
$M_\mu(C_i)\longrightarrow M_\mu(C)$ as $C_i\longrightarrow C$   \\
\end{center}
\caption{\label{min-axioms}
Minkowski functionals are unique descriptors of morphology under the
requirements of Motion Invariance (A), Additivity (B) and Continuity
(C).}
\end{figure}

Let us consider the set of points supplied by cluster positions in
three--dimensional space.  We decorate each point with a ball of
radius $r$, thereby creating connections between neighbouring balls.
We now wish to investigate how the global morphology of the union set
of these balls changes with the radius $r$, which is employed as a
(single) diagnostic parameter.  To achieve this, we need quantitative
measures of geometry and topology for bodies in three--dimensional
space.

It seems sensible to request that such measures be motion invariant
valuations of the bodies, i.e.  scalar functionals satisfying
additivity and invariance under rotations and translations, as well as
a continuity requirement (see Figure~\ref{min-axioms}). The theorem
of Hadwiger (Hadwiger 1957) tells us that in three dimensions any
functional satisfying these requirements is a linear combination of
the four Minkowski functionals. In this sense the four Minkowski
functionals supply a complete and unique characterization of global
morphology in three dimensions.  Table \ref{min-3d-geom} displays
their direct relation to morphological quantities known from
differential geometry. For a smooth body $K$, they are given as
surface integrals of functions of the principal curvature radii $R_1$,
$R_2$:
\begin{equation}
\begin{array}{rcl}
A        &=& 
\displaystyle\int_{\partial{K}}\kern-2ex {\rm d}A, \medskip\\
H        &=& 
\frac{1}{2} \displaystyle\int_{\partial{K}}\kern-2ex {\rm d}A 
\left(\frac{1}{R_1}+\frac{1}{R_2}\right), \medskip\\
\chi &=& 
\frac{1}{4\pi} \displaystyle\int_{\partial{K}}\kern-2ex {\rm d}A 
\frac{1}{R_1R_2}.
\end{array}
\end{equation}
(for a rigorous derivation of these relations see Mecke, Buchert \&
Wagner 1994).  A good overview and extensive references on integral
geometry are given in an article by Weil (1983) or in the recent book
by Schneider (1993).  For further details on Minkowski functionals in
the cosmological context see Mecke (1994), Buchert (1995), Platz\"oder
\& Buchert (1995), Schmalzing, Kerscher \& Buchert (1995) and
especially Mecke, Buchert \& Wagner (1994).

\subsection{Boundary correction}

\begin{figure}
\begin{center}
\epsfxsize=7cm
\begin{minipage}{\epsfxsize}
\epsffile{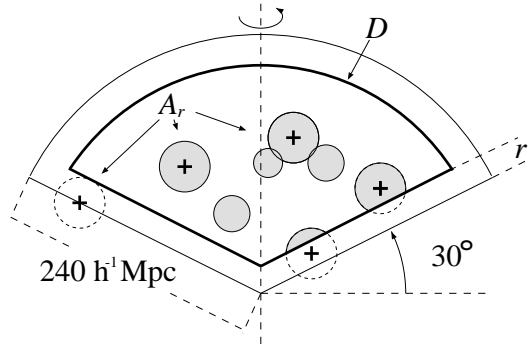}
\end{minipage}
\end{center}
\caption{\label{min-geometry}
Two--dimensional cut through the geometry of one part.  The
shaded area is the set $A_r \cap D$, in this case $D$ is the overall
sample geometry (thin line) shrinked by $r$.}
\end{figure}

For Minkowski functionals there is a concise way of dealing with
boundaries (Mecke \& Wagner 1991). Let $D$ be the window (the sample
geometry) through which we look at $N$ clusters.
$A_r=\bigcup_{i=1}^NB_r(i)$ is the union of balls $B_r(i)$ of
radius $r$ centered on the \mbox{$i$--th} cluster, respectively. In
order to obtain a precise quantitative measure of the boundary
contribution we have to calculate the Minkowski functionals
$M_\mu(A_r\cap{D})$ of the intersection of the union of all balls with
the window, and the Minkowski functionals $M_\mu(D)$ of the window
itself, as illustrated in Figure~\ref{min-geometry}. The quantities
$M_\mu(A_r\cap{D})$ are well suited for the analysis of redshift
catalogues but we can go further.

Following Mecke \& Wagner (1991) and Schmalzing, Kerscher \& Buchert
(1995), we can extract the volume densities of the Minkowski
functionals $m_\mu(A_r)$ from the catalogue if the window contains a
fair sample of an ergodic and stationary point process. The boundary
contribution of the window is then completely removed by applying the
following recursive formula:\footnote{we use the convention
$\sum_{n=i}^j x_n = 0$ for $j<i$}
\begin{equation}
\begin{array}{l}
m_\mu(A_r) = \\
\displaystyle\qquad
\frac{M_\mu(A_r \cap D)}{M_0(D)} - 
\sum_{\nu=0}^{\mu-1} 
\left(\!\!\!\begin{array}{c}{\mu}\\{\nu}\end{array}\!\!\!\right) 
m_\nu(A_r) \frac{M_{\mu-\nu}(D)}{M_0(D)}.
\end{array}
\end{equation}\label{min-deconv}
Fava \& Santal\'o (1979) give a mathematically rigorous derivation of
this formula.
In the case of periodic boundary conditions we have $M_{\nu}(D) = 0$
for \mbox{$\nu \in \{1,2,3\}$} and therefore $m_\mu(A_r)= M_\mu(A_r \cap D) /
M_0(D)$ for \mbox{$\mu\in\{0,1,2,3\}$}.  Generally, for $\mu=0$, the
formula is trivial. We get for the volume density $m_0(A_r)$ (and for
the void probability function, see below)
\begin{equation}
m_0(A_r) = \frac{M_0(A_r \cap D)}{M_0(D)}\;.
\end{equation}
Thus, for the removal of boundary contributions of the Euler
characteristic (and hence the genus, see below) in a window $D$ with
arbitrary non--periodic boundaries one needs to know {\em all} the
Minkowski functionals. Coles, Davies \& Pearson (1996) proposed a
method for calculating the genus of isodensity surfaces based on the
Morse theorem. Their boundary correction is exact for periodic cubes.

\subsection{Connection to other statistics}
\label{min-other-stat}

The most notable Minkowski functional is the Euler characteristic
$\chi$ whose investigation has become a standard method of cosmology
known as genus--statistics (see e.g. Gott, Weinberg \& Melott 1987,
Melott 1990). In three dimensions, for a single body $B$ the Euler
characteristic $\chi(B)$ is related to the genus $g$ of the surface
$\partial B$ by
\begin{equation}
\chi(B) = \frac{1}{2}\chi(\partial B) = \frac{1}{2}(1-g).
\end{equation}
The genus is usually calculated from the integral Gaussian curvature
of smoothed isodensity surfaces. The construction of the isodensity
surface involves two para\-meters, the smoothing length and the density
threshold. In our model we employ the radius as a single diagnostic
scale parameter.
For simulation data, with periodic
boundaries, the construction of the density field is straightforward;
for redshift surveys one has to rely on boundary corrections (Rhoads,
Gott \& Postman 1994). Since we do not need the density field in our
analysis we are not concerned with these additional (empirical)
corrections.

Another well known statistic is the void probability $P_0$ (see White
1979) which is directly related to the volume density $m_0$ according to
\begin{equation}
1-P_0(r) = m_0(A_r).
\end{equation}
Similar to the connection of the void probability with the hierarchy
of correlation functions (see Stratonovich 1963 and White 1979), one
finds analogous expressions for all Minkowski functionals (Mecke 1994,
Mecke, Buchert \& Wagner 1994).

There is a rule of thumb stating that the first zero of the Euler
characteristic serves as an estimate for the percolation threshold
(Mecke \& Wagner 1991).

The dependence of the volume $M_0(A_r)$ on the radius $r$ allows the
calculation of the Minkowski--Bouligand dimension $D_M$
\begin{equation}
D_M = 3 - \lim_{r \rightarrow 0} \frac{\log(M_0(A_r))}{\log(r)}
\end{equation}
which is equal to the capacity (box counting) dimension, giving an
upper limit to the Hausdorff dimension (see Falconer 1990).  As in the
case of the ordinary box--counting method, any estimate of $D_M$ based
on a finite number of points is affected by discreteness problems, as
discussed by Borgani et al. (1993) and Dubrulle \& Lachi\`eze-Rey
(1994).

\subsection{Analysis of the Abell/ACO catalogue}
\label{min-dat}

\begin{figure}
\begin{center}
\epsfxsize=3in
\begin{minipage}{\epsfxsize}
\epsffile{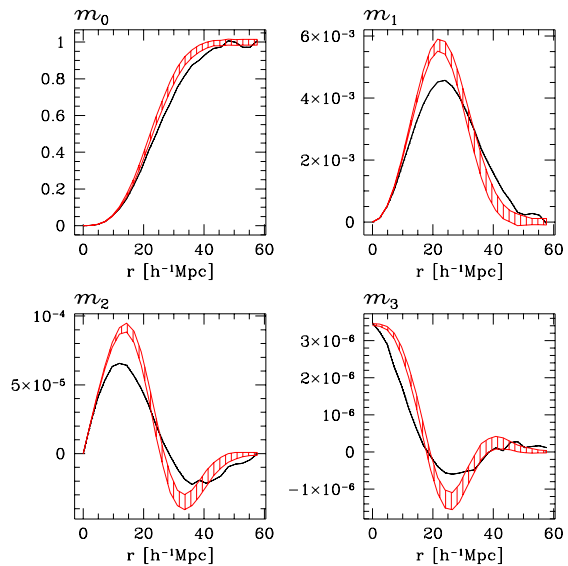}
\end{minipage}
\caption{\label{min-abell}
Densities of the Minkowski functionals for the Abell/ACO (average of
northern and southern parts) and a Poisson process (shaded area)
with the same number density. The shaded area gives the statistical
variance of the Poisson process calculated from 100 different
realizations.}
\end{center}
\end{figure}

In order to calculate the densities of Minkowski functionals from the
Abell/ACO sample, we analyzed the northern and southern part
separately. Then we calculated the average between these two parts;
since they do not intersect, this gives the same results as analyzing
both parts together. In Figure \ref{min-geometry} we sketch the geometry
of one part. At each radius $r$, the walls of the original window are
shrunk by the radius $r$. All clusters within the original window
contribute, but clusters lying outside the original window do not. We can
then apply the boundary removal described in Subsection \ref{min-deconv}
for each radius separately to recover the densities $m_\mu(A_r)$ of the
Minkowski functionals. Figures \ref{min-abell}, \ref{min-cdm-tcd-abell}
and \ref{min-lcd-bsi-abell} display these densities for the various
processes described above. 

According to their definition, the densities of the Minkowski functionals 
are expressed in the following units:
\[
\begin{array}{ll}
[m_0] = 1, &[m_1] = (\hm)^{-1}, \\ 
{[}m_2] = (\hm)^{-2}, & [m_3] = (\hm)^{-3} .
\end{array}
\]

In Figure \ref{min-abell} we plot the densities of the Minkowski
functionals for both the Abell/ACO sample (solid lines) and a Poisson
process (shaded area). Although the mean values for the Poisson
process are known analytically (Mecke \& Wagner 1991), in Figure
\ref{min-abell} we preferred to compute numerically the mean values
and the standard errors of the Minkowski functionals for 100
realizations of such a Poisson process within the sample geometry.

The most prominent feature of all four Minkowski functionals are the
broader extrema for the Abell/ACO data as compared to the results for
the Poisson process. This is a first indication for enhanced
clustering. Let us now look at each functional in detail.

The Minkowski functional $m_0$ measures the density of the covered
volume.  On scales between $25\hm$ and $40\hm$, $m_0$ as a function of
$r$ lies slightly below the Poisson data. The volume density is lower
because of the clumping of clusters on those scales. It is remarkable
that this behaviour occurs in the spatial region where clusters are
weakly clumped as measured by the conventional two--point
autocorrelation function (the region where $0 < \xi(r) < 1$).  The
Minkowski functionals involve correlation functions of every order
(see Section \ref{min-other-stat}) and therefore are more sensitive to
enhanced clumping.

The Minkowski functional $m_1$ measures the surface density of the
coverage. It has a maximum at about $20\hm$ both for the Poisson
sample and for the cluster data. This maximum is due to the granular
structure of the union set on the relevant scales. At the same scales,
we find the maximum deviation from the Poisson distribution. The lower
values of the cluster data $m_1$ with respect to the Poisson samples
is again the signature for the presence of significant clumping of
clusters at these scales. The functional $m_1$ shows also a positive
deviation from the Poisson samples on scales $(35$\dots$50)\hm$ where
more coherent structures form in the union set than in the Poisson
samples, keeping the surface density larger.

The Minkowki functionals $m_2$ and $m_3$ characterize in more detail
the kind of spatial coverage provided by the union set of balls in the
data sample. The density of the total mean curvature $m_2$ of the data
reaches a maximum at about $10\hm$ produced by the dominance of convex
(positive $m_2$) structures.  The density $m_2$ at the maximum is
reduced with respect to the Poisson sample to about 70\% (or more than
3 standard deviations).  The integral mean curvature $m_2$ has a zero
at a scale of $25\hm$ (almost the scale of maximum of $m_1$)
corresponding to the turning--point between structures with mainly
convex and concave boundaries (negative $m_2$). Significant deviations
from the Poisson distribution occur between this turning point and $40
\hm$ due to the smaller mean curvature of the union set of the data,
probably caused by the interconnection of the void regions in the
cluster distribution.

The density of the Euler characteristic $m_3$ describes the global
topology of the cluster distribution. On small scales all balls are
separated. Therefore, each ball gives a contribution of unity to the
Euler characteristic and $m_3$ is proportional to the cluster number
density. As the radius increases, more and more balls overlap and
$m_3$ decreases. At a scale of about $20\hm$ it drops below zero due
to the emergence of tunnels in the union set (a double torus has
$\chi=-1$).  The positive maximum for the Poisson process at scales
$\simeq 40\hm$ is the signature for the presence of cavities.  The
nearly linear decrease of the Euler characteristic for the Abell/ACO
sample indicates strong clustering on scales $\mincir15\hm$.  This
confirms the results of Bahcall (1988), stating that superclusters
consist mainly of pairs or triples of strongly correlated (rich)
clusters.  The lack of a significant positive maximum after the
minimum shows that only a few cavities form.  This suggests on such
scales a support dimension for the distribution of clusters of less
than three, (see also Borgani\etal 1993);
note that the formation of cavities resulting in positive
contributions to the Euler characteristic is not possible for a
support dimension $\le 2$.  The presence of voids on scales of $30$ to
$45\hm$ is shown by the enhanced surface area $m_1$ and the reduced
integral mean curvature $m_2$, while on these scales the Euler
characteristic $m_3$ is approximately zero.  This does not confirm a
pure shell model, proposed by Bahcall (1988) and used by Mecke,
Buchert \& Wagner (1994) as the ``double Poisson process''; rather,
the shape of the cluster distribution is more likely described by a
mixture of cavities and tunnels.  However, since the scatter in the
Minkowski functionals on scales above $45\hm$ is quite high, this last
interpretation remains to be confirmed by larger future data sets.

\subsection{Analysis of mock catalogues}

\begin{figure}
\epsfxsize=3in
\begin{center}
\begin{minipage}{\epsfxsize}\epsffile{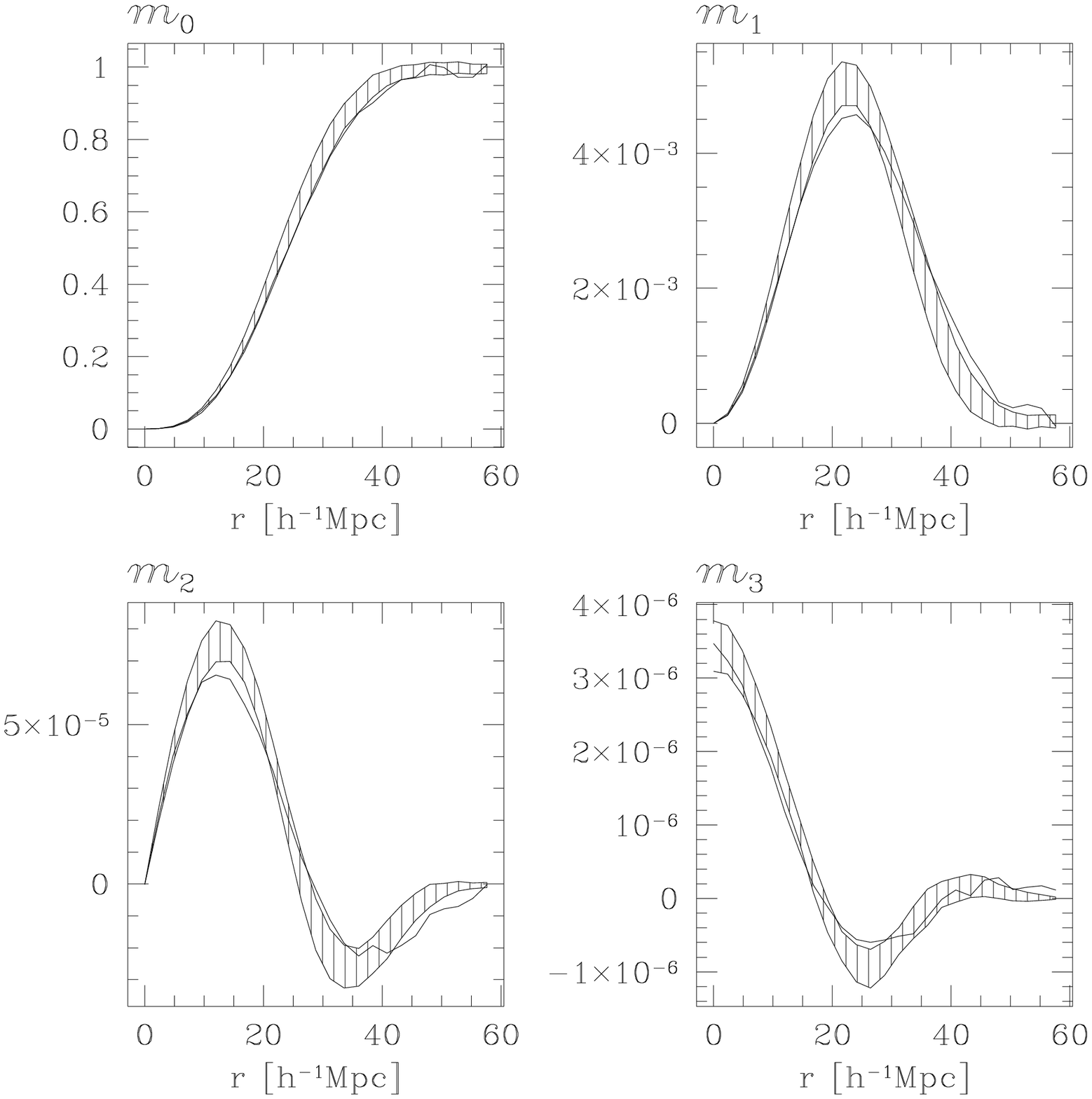}\end{minipage}
\begin{minipage}{\epsfxsize}\epsffile{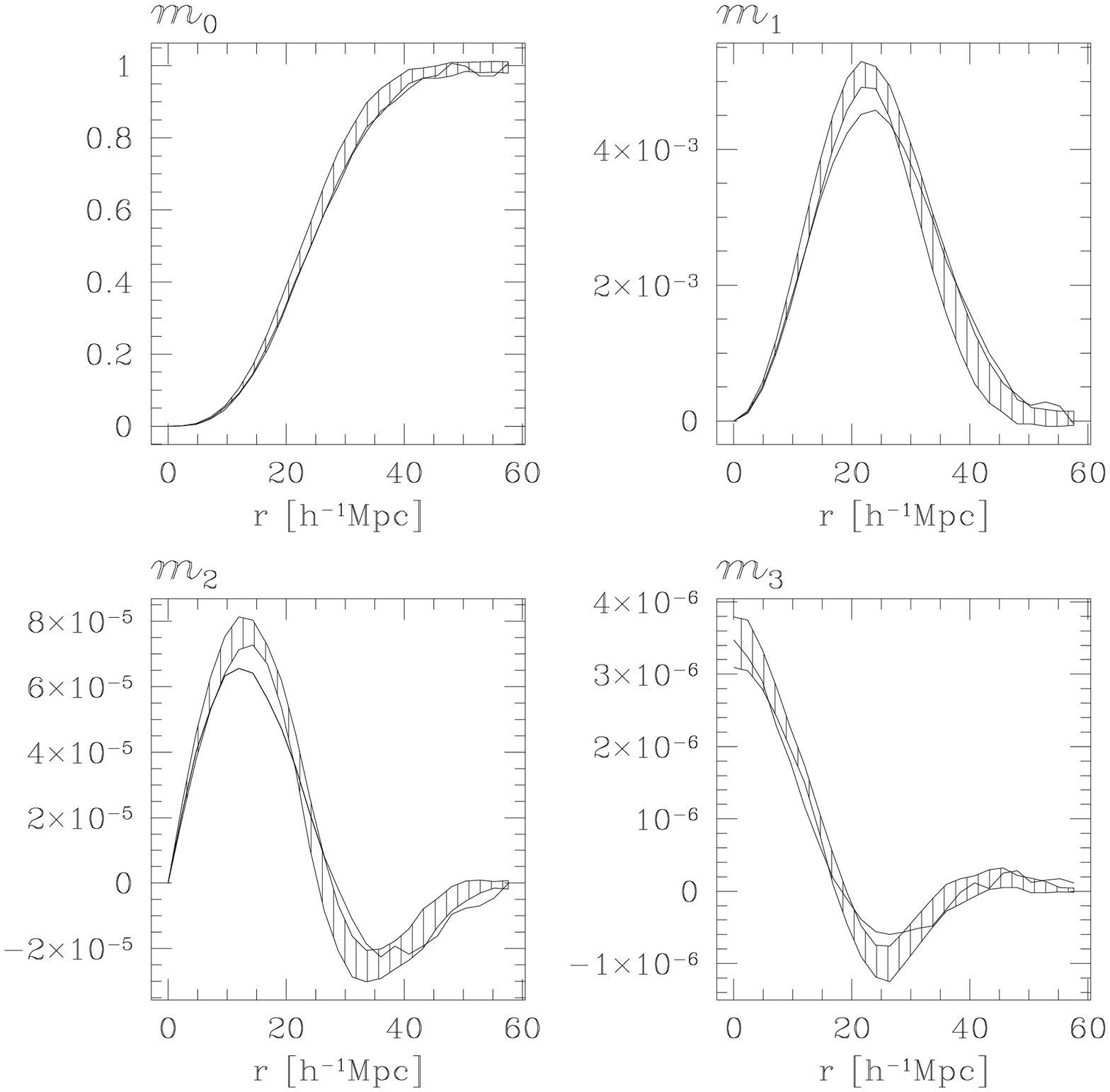}\end{minipage}
\end{center}
\caption{\label{min-cdm-tcd-abell}
Densities of the Minkowski functionals for the Abell/ACO (solid line
in both panels) compared to the SCDM (shaded area in top panel), and
the TCDM (shaded area in bottom panel).  The shaded area gives
1$\sigma$--errorbars of the variance among different realizations
including cosmic variance, as explained in the text.}
\end{figure}

\begin{figure}
\epsfxsize=3in
\begin{center}
\begin{minipage}{\epsfxsize}\epsffile{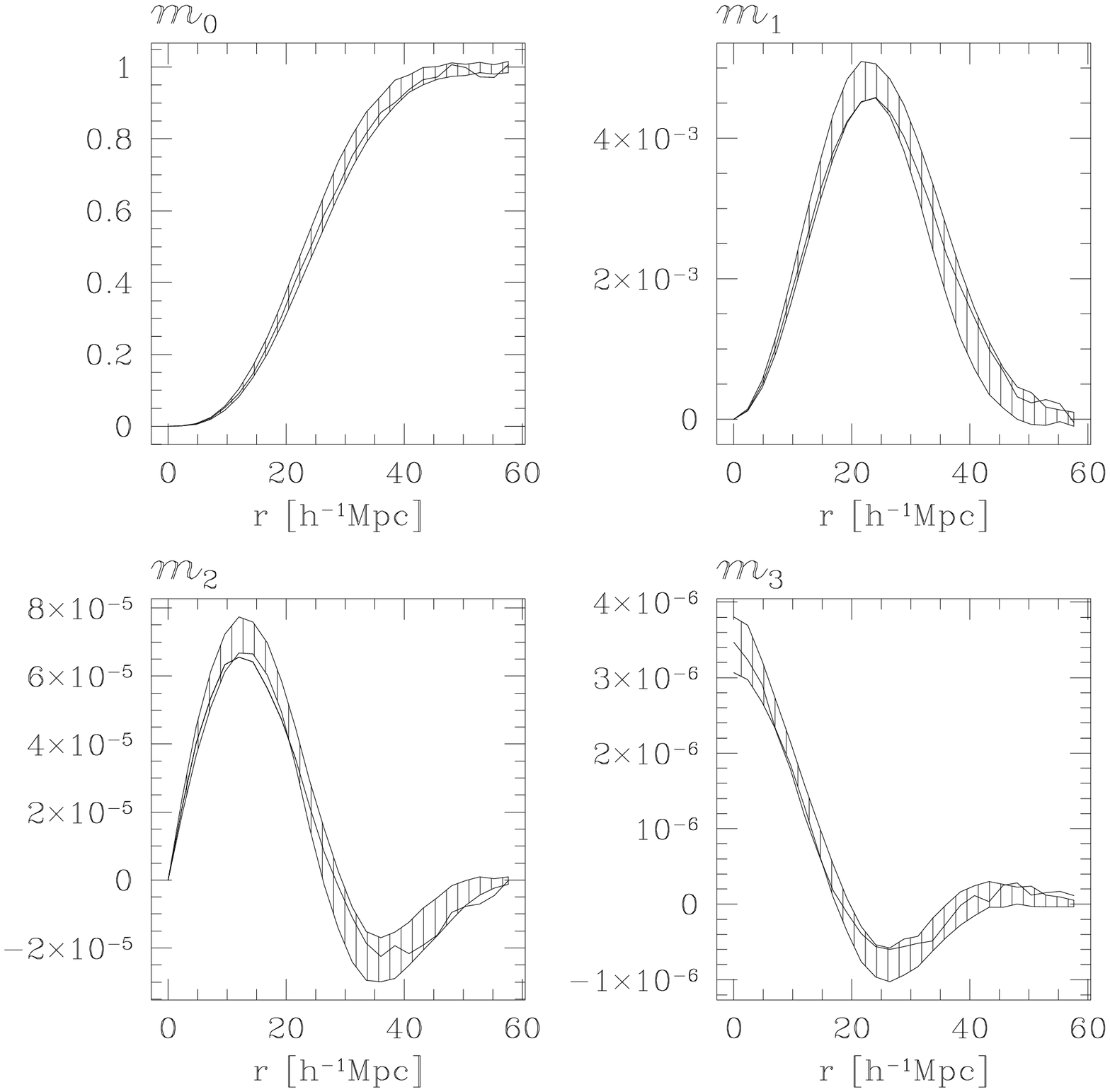}\end{minipage}
\begin{minipage}{\epsfxsize}\epsffile{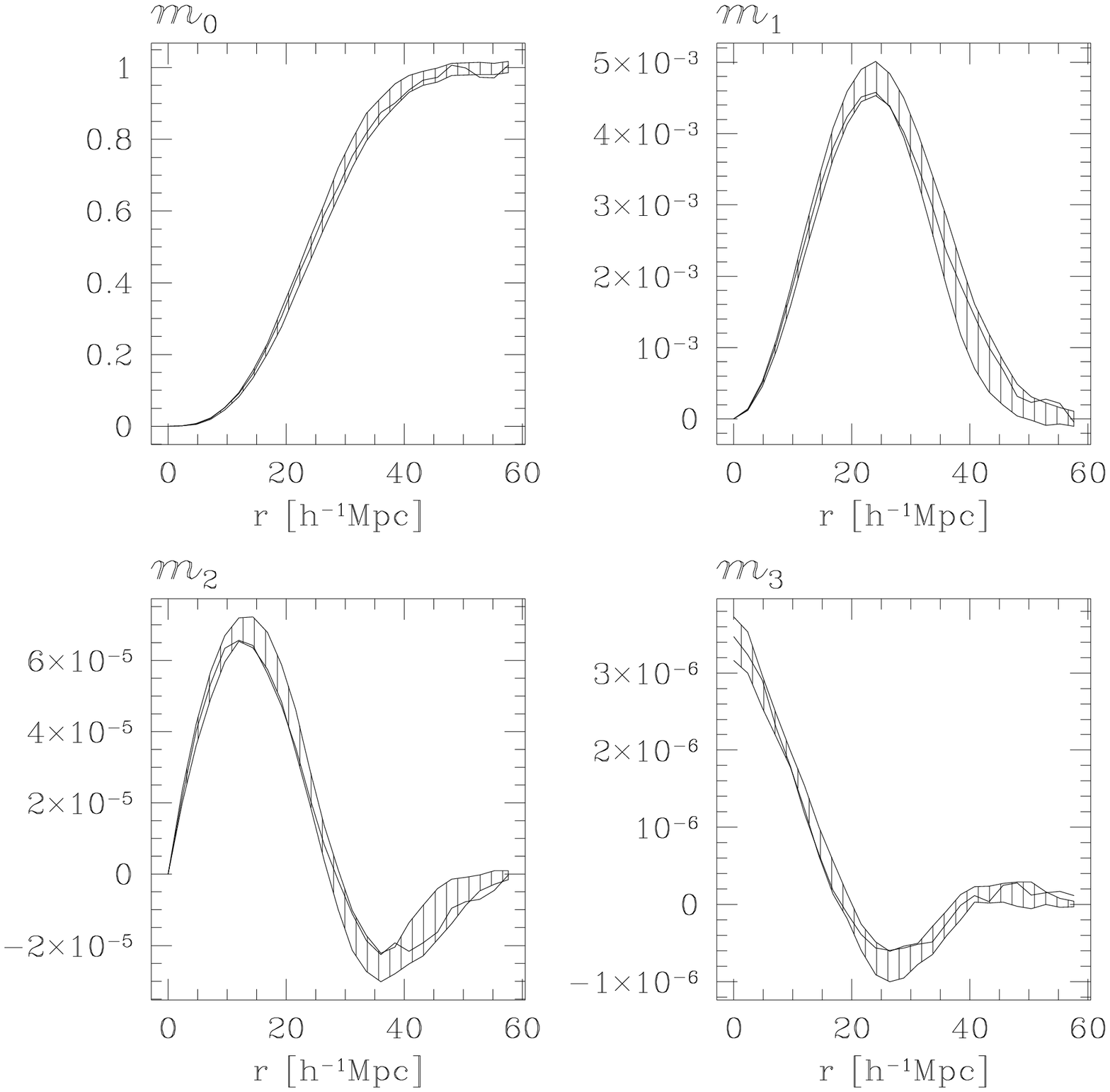}\end{minipage}
\end{center}
\caption{\label{min-lcd-bsi-abell}
Densities of the Minkowski functionals for the Abell/ACO (solid line
in both panels) compared to the $\Lambda$CDM (shaded area in top
panel), and the BSI (shaded area in bottom panel).  The shaded area
gives 1$\sigma$--errorbars of the variance among different
realizations including cosmic variance, as explained in the text.}
\end{figure}

Since the construction of mock samples by the algorithm described in
Section~\ref{mock-sect} takes into account all selection effects of
the Abell/ACO catalogue, we can analyze them by using the same
procedure as for the real data. The considerable deviations of the
mock samples' number density from the mean cluster number density, as
defined within the whole simulation box, are clearly seen in the
scatter of the density of the Euler characteristic $m_3$ at the
smallest radii, which is proportional to the number density (see
Figures {}\ref{min-cdm-tcd-abell} and {}\ref{min-lcd-bsi-abell}).
This suggests that even a sample as large as the Abell/ACO catalogue
is not a `fair sample' as discussed by Buchert and Mart\'{\i}nez
(1993), since it is still significantly influenced by cosmic variance.

A comparison of the four SCDM (or the three $\Lambda$CDM) realizations
shows that this scatter in the number density is the main source for
the variance seen in the Minkowski functionals.  The statistical
variance introduced by the scatter of different realizations of the
initial density field is always negligible if compared to the
observer--to--observer scatter in each box, associated with the number
density fluctuations between different mock samples.  In this sense
the error bands shown in the plots include cosmic variance; only the
`robustness' (a result of the additivity property) and the high
significance of the Minkowski functionals still allow for discrimination
between the different Dark Matter cosmologies.

In Figure {}\ref{min-cdm-tcd-abell} we compare the densities of the
functionals for SCDM and TCDM models to the Abell/ACO ones. Both
models show too little clustering on small scales, as it is clearly
seen by the enhanced maxima of the surface area $m_1$ and the integral
mean curvature $m_2$, as well as in the flatter decrease of the Euler
characteristic $m_3$. Additionally, the higher volume $m_0$ indicates
weak clumping also on large scales.

Figure {}\ref{min-lcd-bsi-abell} displays the same quantities as
Figure {}\ref{min-cdm-tcd-abell} but for the $\Lambda$CDM and BSI
models. Although small deviations from the Abell/ACO results are still
present, again pointing towards weak clumping, both models perform
much better than SCDM and TCDM.

\begin{figure}
\epsfxsize=3in
\begin{center}
\begin{minipage}{\epsfxsize}
\epsffile{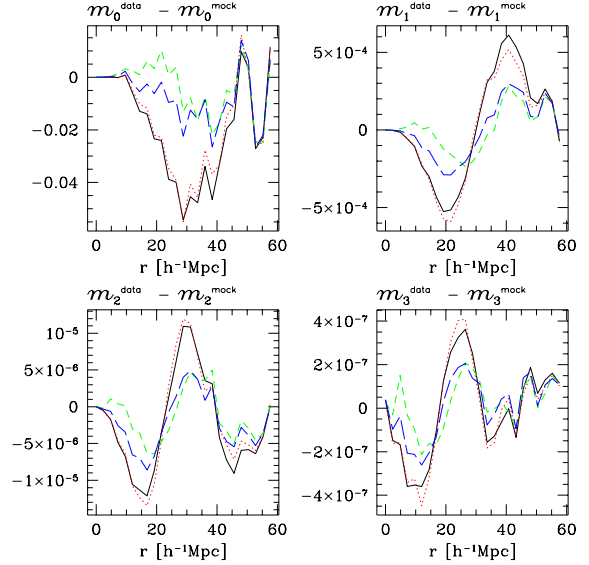}
\end{minipage}
\end{center}
\caption{\label{min-differences}
Differences of the densities of Minkowski functionals for the SCDM
(solid line), TCDM (dotted), $\Lambda$CDM (long dashed), and the BSI
(short dashed).}
\end{figure}

This is also confirmed by the plot in \mbox{Figure
\ref{min-differences}}, where we show the differences between the
functionals of the Abell/ACO data, $m_\mu^{\rm data}$, and the
functionals of the mock samples, $m_\mu^{\rm mock}$.  Again we see
that $\Lambda$CDM and BSI describe the cluster distribution much
better than SCDM and TCDM.
At scales above $45 \hm$ the differences between the Minkowski
functionals of the Abell/ACO and the mock samples are dominated by
random fluctuations in the Abell/ACO.

\begin{figure}
\begin{center}
\epsfxsize=3in
\begin{minipage}{\epsfxsize}
\epsffile{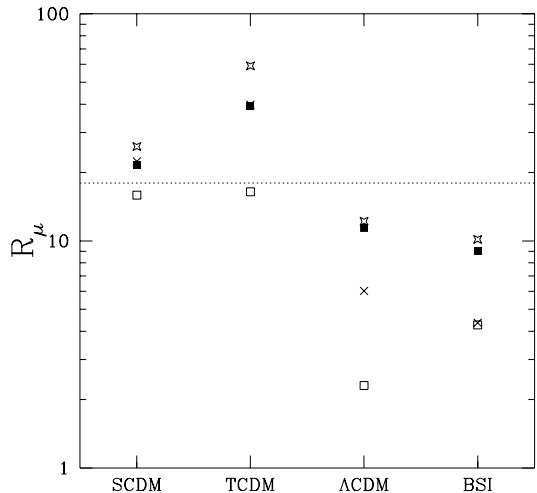}
\end{minipage}
\end{center}
\caption{\label{min-risk}
The risk values for simulations versus Abell/ACO for the volume
measure $R_0$ (empty boxes), for the surface density $R_1$ (crosses), for
the mean curvature $R_2$ (stars), and for the Euler characteristic
$R_3$ (filled boxes).}
\end{figure}

In order to quantify the statistical significance of these deviations
we calculate the risks $R_\mu$, with an error weighted quadratic cost
function, for all functionals of all models versus those of the
Abell/ACO sample. This risk is equal to the $\chi^2$ distance used in
maximum likelihood analysis (see e.g.~Frieden 1991). For $N_r$ radii
it is given by
\begin{equation}
R_\mu = \sum_{i=1}^{N_r} \frac{(m_\mu^{\rm mock}(r_i)-
m_\mu^{\rm data}(r_i))^2}{\sigma_\mu^{\rm mock}(r_i)^2},
\end{equation}
where $\sigma_\mu^{\rm mock}(r)$ is the standard error of the
Minkowski functionals at radius $r$ over several mock catalogues,
\begin{equation}
\sigma_\mu^{\rm mock}(r) = \sqrt{\langle m_\mu^{\rm mock}(r)^2\rangle-
\langle m_\mu^{\rm mock}(r)\rangle^2}.
\end{equation}
In Figure~\ref{min-risk} we show the $R_\mu$ values for the four
considered DM models.  After excluding $r=0$ and $r> 45 \hm$, we limit
the analysis to $N_r=18$ radial bins.  Therefore, a constant departure
of 1$\sigma$ should result in a risk $R_\mu=18$, which is shown for
reference in Figure~\ref{min-risk} as a dotted line.
SCDM shows a significant departure from Abell/ACO, as a result of the
reduced clustering. The differences of TCDM are even larger than those
of SCDM (cf.~also Figure \ref{min-differences}). The higher risk
values for TCDM may be caused by underestimating $\sigma_\mu$ (we
recall that we analyzed only one TCDM simulation, i.e.~eight mock
catalogues, but four SCDM simulation, i.e.~32 mock catalogues).
As for $\Lambda$CDM and BSI, they are confirmed to fit the data much
better; the difference with respect to Abell/ACO data is below the
1$\sigma$ level for all the functionals, but showing also a tendency
towards too weak clumping.
The volume density (i.e.~the void probability) alone is not
sufficiently discriminating (see also Figures \ref{min-cdm-tcd-abell},
\ref{min-lcd-bsi-abell}), while the integral mean curvature $m_2$ is
even more selective than the Euler characteristic $m_3$.

\section{Summary and Conclusion}

We have calculated scale--dependent Minkowski functionals for spatial
patterns induced by the point set of an Abell/ACO cluster redshift
sample and compared these functionals with those obtained from a
number of mock cluster catalogues obtained from different cosmological
models. We find significant deviations in the morphological features
of the observed cluster clumping from the Poisson distribution on
scales $(15$\ldots$50)\hm$, which have not been explored previously in
such detail. In particular, in the range $(15$\ldots$45)\hm$ the results
indicate a clumping of clusters on a support with dimension less than
three (see also Borgani et al.~1994). On scales $(25$\ldots$45)\hm$ the
behavior of the Minkowski functionals $m_1$ (area), $m_2$ (integral
mean curvature) and $m_3$ (Euler characteristic) may be interpreted as
a cluster aggregation exhibiting cavities and interconnected tunnels
rather than isolated voids.
Our results from the mock catalogues agree with the expectation that
the SCDM and TCDM models do not describe the degree of cluster
clumping as inferred from the data in the scale range
$(15$\ldots$45)\hm$.  Moreover the formation of cavity aggregates is
not reproduced convincingly. On the other hand, the $\Lambda$CDM and
the BSI models provide a reasonably good although not perfect
description of the data.

The main source of uncertainty in the analysis arises from the
fluctuations in the number density of objects among the mock
catalogues. The deviations of the model results from the data are on
the 1$\sigma$ level. Nevertheless we think that the results are
reliable, since they are inferred from the behaviour of our
morphological measures over an extended range of the diagnostic length
scale parameter $r$.

The family of Minkowski functionals focuses on global features of
spatial patterns and includes, besides a topological descriptor for the
connectivity, also geometrical measures for the size and shape. Thus
the present approach provides a unifying frame for the analysis of
cosmic structures which comprises the void probability function as well
as the genus statistics; our method is introduced not to replace but to
complement traditional tools such as low--order correlation functions.

\section*{}
The numerical code used for calculating the Minkowski functionals is
available and can be obtained by sending e--mail to \mbox{\tt
buchert@stat.physik.uni-muenchen.de}.

\section*{Acknowledgements}

MK and TB acknowledge support from the ``Sonderforschungsbereich SFB
375 f\"ur Astroteilchenphysik der Deutschen
Forschungsgemeinschaft''. JR is supported by the Deutsche
Forschungsgemeinschaft through grant Go563/5-2. SB, JS and MK thank the
Astrophysical Institute of Potsdam for hospitality while preparing
part of this work.

\end{document}